\newcommand\aap{{A\&A\,}}%
\documentclass[graybox, natbib, footinfo]{svmult}

\usepackage{mathptmx}       
\usepackage{helvet}         
\usepackage{courier}        
\usepackage{type1cm}        
\usepackage{makeidx}         
\usepackage{graphicx}        
\usepackage{multicol}        
\usepackage[bottom]{footmisc}

\makeindex             

\begin{document}

\title*{The red giants in NGC 6633 as seen with CoRoT, HARPS and SOPHIE}
\titlerunning{NGC 6633 red giants} 
\author{Ennio Poretti, Philippe Mathias, Caroline Barban, Frederic Baudin, 
Andrea Miglio, Josefina Montalb\'an, Thierry Morel, and Benoit Mosser}
\authorrunning{E. Poretti et al.} 
\institute{Ennio Poretti \at INAF-Osservatorio Astronomico di Brera, Via E. Bianchi 46, 23807
Merate, Italy, \email{ennio.poretti@brera.inaf.it}
\and Philippe Mathias \at 
CNRS, IRAP, 57 Avenue d'Azereix, BP 826, 65008 Tarbes, France
\and Caroline Barban, Frederic Baudin, Benoit Mosser
\at LESIA, CNRS, Universit\'e Pierre et Marie Curie, Universit\'e
 Denis Diderot, Observatoire de Paris, 92195 Meudon Cedex, France
\and Thierry Morel and Josefina Montalb\'an \at Institut d'Astrophysique et de G\'eophysique, Universit\'e
de Li\`ege, Li\`ege, Belgium
\and Andrea Miglio \at School of Physics and Astronomy, University of Birmingham,
Edgbaston, Birmingham, B15 2TT, UK
}
%
%
\maketitle

\abstract{The open cluster NGC 6633 was observed with CoRoT in 2011 and simultaneous high-resolution
spectroscopy was obtained with the SOPHIE and HARPS spectrographs. One of the four targets was not
found  to be a cluster member. For all stars we provide estimates of the seismic and
spectroscopic parameters.  }
\vskip 1.0truecm


The CoRoT satellite (Convection, Rotation and planetary Transits, \citealt{Baglin06}) 
observed the open cluster NGC 6633 in two long runs allocated 
from April 2011 to September 2011 (LRc07 and LRc08 in the CoRoT schedule).
The  red giants HD 170031 ($V$=8.20), HD 170231 ($V$=8.69), HD 170053 ($V$=7.30),
HD 170174 ($V$=8.31)
and the  B8-star HD 170200 ($V$=5.70) were the five stars selected to be
observed in the asteroseismic channel. 
The Seismologic Ground-Based Working Group  considered very challenging
to perform simultaneous radial velocity measurements to measure the 
amplitude of the radial velocity variations of the solar-like oscillations 
occurring in red giants.

\begin{figure}[t]
\sidecaption
\includegraphics[scale=.30]{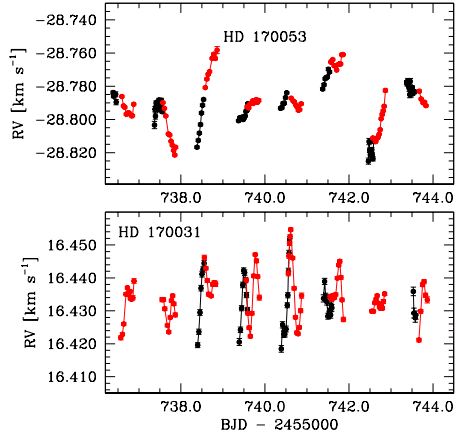}
\caption{Radial velocity measurements of HD 170031 and HD 170053 obtained with SOPHIE (black circles)
and HARPS (red circles).}
\label{curve}
\end{figure}

The spectroscopic runs with HARPS at European Southern Observatory were scheduled 
from June 23 to July 3, 2011 and from 15 to 20 July, 2011. 
The runs with SOPHIE at Observatoire Haute Provence  from May 26 to June 6, 2011, and
from June 20 to July 1, 2011.  
Due to 
long exposure time requested to provide the
necessary accuracies, only HD 170031 and HD 170053 could be intensively observed at both observatories.
Figure~\ref{curve}  shows the radial velocity curves obtained from June~23 to July~1, 2011.
Note that error bars are reported, but often not noticeable since they have the same 
size of the points.

The double-site campaign clearly put in evidence the multiperiodic behaviours of
the radial velocity variation in both the well-sampled red giants.
Peak-to-peak
amplitudes are 60~m\,s$^{-1}$ for HD 170053 and 40~m\,s$^{-1}$ for HD 170031. 
The uncertainty on a single measurement is $\pm$1~m\,s$^{-1}$.
It also resulted that HD 170053, HD 170231, and HD 170174 have the same mean radial velocity
(around $-$28~km\,s$^{-1}$), while HD 170031 shows a completely different value 
(Fig.~\ref{curve}).
The NGC 6633 radial velocity is $-$25.43~km\,s$^{-1}$ \citep{Kharchenko05} and 
therefore we can conclude that, unlike the other three red giants, 
HD 170031 is not a cluster member.

\begin{table}
\caption{Physical and seismic parameters of the red giants in NGC 6633}
\label{para}       
\begin{tabular}{p{2cm}p{1cm}p{1cm}p{1cm}p{1cm}p{1cm}p{1cm}p{1cm}}
\hline\noalign{\smallskip}
Star & $\nu_{\rm max}$ &  $\Delta\nu$ &  T$_{\rm eff}$ & $\log$ g & [Fe/H] & $\xi$\\
     & [$\mu$Hz] & [$\mu$Hz] &  [K] & [cgs] & &  [km\,s$^{-1}$] \\
\hline\noalign{\smallskip}
HD 170174 & 44.56  & 4.17 &  5055 & 2.56 & --0.07 & 1.58 \\
HD 170053 &  9.18  & 1.09 &  4290 & 1.85 & --0.03 & 1.68 \\
HD 170231 & 66     & 5.33 &  5175 & 2.74 & --0.03 & 1.49 \\
HD 170031 & 39     & 3.87 &  4515 & 2.46 &  +0.04 & 1.41 \\
\noalign{\smallskip}\hline\noalign{\smallskip}
\end{tabular}
\end{table}

\begin{figure}
\sidecaption
\includegraphics[scale=.35]{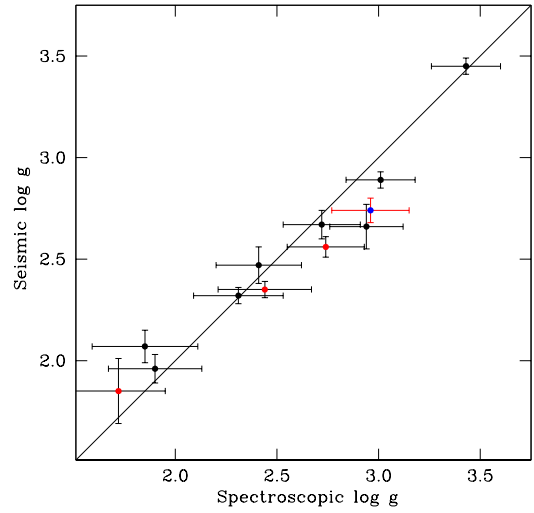}
\caption{Agreement between $\log$\,g values obtained from seismic and spectroscopic analyses.
Red circles: red giants belonging to NGC 6633; blue circle: HD~170031; black circles: galactic
red giants.  }
\label{logg}
\end{figure}

The HARPS spectra were also used to accurately estimate the atmospheric  parameters
(effective temperature T$_{\rm eff}$, surface gravity $\log$\,g, metallicity
[Fe/H], microturbulent velocity $\xi$) 
and the abundances of 16 chemical species in a self-consistent manner \citep{Morel14},
for both  NGC~6633 targets  and other galactic red giants.
Moreover, the extensive photometric CoRoT timeseries supplied 
reliable estimates both for the frequency of the maximum oscillation power
($\nu_{max}$) and for the large separation ($\Delta\nu$). 
The atmospheric parameters could be obtained from the seismic analyses, too
(Table~\ref{para}).
The agreement is excellent, as Fig.~\ref{logg} shows  for the surface gravities.
In particular, note the smaller error bars of the seismic values.



This study shows that spectroscopic and photometric data can be
combined to describe the pulsational
scenario of red giants and, in turn, make more complete
models of their atmospheres.


\begin{acknowledgement}
The CoRoT space mission, launched on 2006 December 27, was developed 
and is operated by the CNES, with participation of the Science Programs 
of ESA, ESA's RSSD, Austria, Belgium, Brazil, Germany and Spain.
This work is based on observations collected at La Silla Observatory, ESO (Chile) 
under program LP185.D-0056.
\end{acknowledgement}

%


\begin{thebibliography}{3}
\expandafter\ifx\csname natexlab\endcsname\relax\def\natexlab#1{#1}\fi

\bibitem[{{Baglin} {et~al.}(2006){Baglin}, {Auvergne}, {Barge}, {Deleuil},
  {Catala}, {Michel}, {Weiss}, \& {COROT Team}}]{Baglin06}
{Baglin}, A., {Auvergne}, M., {Barge}, P., {et~al.} 2006, in ESA Special
  Publication, Vol. 1306, ESA Special Publication, ed. M.~{Fridlund},
  A.~{Baglin}, J.~{Lochard}, \& L.~{Conroy}, 33

\bibitem[{{Kharchenko} {et~al.}(2005){Kharchenko}, {Piskunov}, {R{\"o}ser},
  {Schilbach}, \& {Scholz}}]{Kharchenko05}
{Kharchenko}, N.~V., {Piskunov}, A.~E., {R{\"o}ser}, S., {Schilbach}, E., \&
  {Scholz}, R.-D. 2005, \aap, 438, 1163

\bibitem[{{Morel} {et~al.}(2014){Morel}, {Miglio}, {Lagarde}, {Montalb{\'a}n},
  {Rainer}, {Poretti}, {Eggenberger}, {Hekker}, {Kallinger}, {Mosser},
  {Valentini}, {Carrier}, {Hareter}, \& {Mantegazza}}]{Morel14}
{Morel}, T., {Miglio}, A., {Lagarde}, N., {et~al.} 2014, \aap, 564, A119

\end{thebibliography}
\end{document}